\documentclass[pra,aps,twocolumn,showpacs]{revtex4}

\usepackage{graphicx}

\newcommand{\rme}{{\rm e}}
\newcommand{\rmi}{{\rm i}}
\newcommand{\rmd}{{\rm d}}

\newcommand{\bra}[1]{\langle#1|}
\newcommand{\ket}[1]{|#1\rangle}
\newcommand{\beq}{\begin{equation}}
\newcommand{\eeq}{\end{equation}}
\newcommand{\bfr}{\mathbf{r}}
\newcommand{\bfk}{\mathbf{k}}
\newcommand{\bfq}{\mathbf{q}}

\newcommand{\bfp}{\mathbf{p}}

\begin{document}
\title{Single Atom Cooling by Superfluid Immersion: A Non-Destructive Method for Qubits}
\author{A. J. Daley}
\author{P. O. Fedichev}
\author{P. Zoller}
\affiliation{Institut f\"{u}r Theoretische Physik, Universit\"{a}t Innsbruck, A-6020 Innsbruck, Austria}
\date{25 August 2003}

\begin{abstract}
We present a scheme to cool the motional state of neutral atoms confined in sites of an optical lattice by immersing the system in a superfluid. The motion of the atoms is damped by the generation of excitations in the superfluid, and under appropriate conditions the internal state of the atom remains unchanged. This scheme can thus be used to cool atoms used to encode a series of entangled qubits non-destructively. Within realisable parameter ranges, the rate of cooling to the ground state is found to be sufficiently large to be useful in experiments. 
\end{abstract}
\pacs{03.67.Lx, 42.50.-p} 

\maketitle

\section{Introduction}

Neutral atoms are one of the most promising candidates as carriers for the storage and manipulation of quantum information \cite{qcompgeneral}. Qubits may be stored in long-lived internal atomic states with very low levels of decoherence, and may be manipulated using interactions between the atoms and external devices (such as lasers) or interactions amongst the atoms themselves. 

An experimental prerequisite for this is the development of techniques to trap single atoms, and there has been much progress over the last five years both in optical traps \cite{opticaltrap,greiner} and in magnetic microtraps \cite{magtrap}. In addition, specific implementation of quantum computing usually requires cooling of atoms to the vibrational ground state of the trap, or at least to the Lamb-Dicke limit. Many techniques have been developed including the widespread use of laser cooling \cite{metcalf}. 

One of the most promising routes to quantum computation with neutral atoms is the use Bose Einstein Condensates (BECs) \cite{becreview} loaded in optical lattices \cite{deutschreview,oltheory}, a system which has been realised in part in a number of recent experiments \cite{greiner,sdol,inguscioexp,arimondoexp}. There are several theoretical proposals for the implementation of quantum logic gates in such systems \cite{calarcoreview,deutsch,briegeloneway,lewenstein,williams}, and the first steps towards the fundamental experimental techniques required for some of these have been recently realised. For example, the recent demonstration of spin-dependent transport in an optical lattice \cite{sdol} makes possible the implementation of a fundamental quantum phase gate by cold controlled collisions \cite{calarcoreview} in which qubits are encoded using two different internal states of the atoms in the optical lattice.

However, most of these proposals require the transport of qubits, which is usually associated with heating of the atomic motion \cite{sdol}. A question then arises as to how that motion may be cooled back to the ground state without changing the internal state of the atoms, and thus destroying the qubits or their entanglement. Laser cooling, for example, is clearly not applicable here as the process of light scattering causes decoherence. The same problem arises in scalable ion trap quantum computing, and there it has been overcome using sympathetic cooling schemes, in which ions used to encode qubits are cooled via a coulomb interaction with either a single ion which is directly laser-cooled \cite{blatt} or another species of ions which are directly laser-cooled \cite{iontrap}. In a different context, sympathetic cooling schemes are also used widely in the field of cold quantum gases, where they have been used to cool different spin states of the same atomic species \cite{scsamespecies}, to cool different Bosonic species \cite{scbosons}, and to cool Fermi gases brought into contact with a BEC \cite{scfermi}.

In this article we consider the sympathetic cooling of a single atom in a harmonic trap in contact with a superfluid. This is readily expanded to the case of many harmonic traps, which is a good approximation for an optical lattice without tunneling.  The motion of the atom is damped by the generation of excitations in the superfluid, and the resulting cooling rates are sufficiently large to be useful experimentally. In addition, decoherence of a qubit encoded on the atoms can be eliminated in this scheme provided that the internal atomic states used to encode the qubit are chosen carefully in order to satisfy particular collisional requirements. 

\section{Overview}
\label{overview}
In this section we give a short summary of the most important results
contained in this article. Derivations and further discussions of these
results follow in the remaining sections.

Our goal is to cool a single trapped atom representing a qubit $|0\rangle
$, $|1\rangle $ without destroying the superposition state of the qubit (or the
entangled state in case of many atoms). Cooling of the atom is achieved by
sympathetic cooling, immersing the atom in a superfluid, which plays the
role of a very cold reservoir. By a proper encoding of the qubit in internal
atomic states, and choice of the atomic level for the superfluid reservoir
(see section \ref{decoherencesupp}) we can ensure that (i) the qubit is not destroyed by opening
collisional channels to unwanted final states, and (ii) the $|0\rangle $ and
$|1\rangle $ states have identical collisional properties with respect to
the collisional interactions with the superfluid, and thus the collisions do not
randomise the relative phases of the qubit.

Cooling is considered within a model in which the atoms are treated as being
trapped in independent 1D Harmonic oscillator potentials with trapping
frequency $\omega$, and interact with the superfluid via a density-density
interaction, generating excitations in the superfluid, which are modeled as
Bogoliubov excitations in a weakly interacting Bose gas (section
\ref{model}%
) and have momentum $\hbar \mathbf{q}$ and energy $\varepsilon _{q}$. In
discussing this cooling process we can restrict ourselves to a single
component of the qubit $|0\rangle$ (or $|1\rangle$).
 
\begin{figure}[htb]
\includegraphics[width=7.5cm]{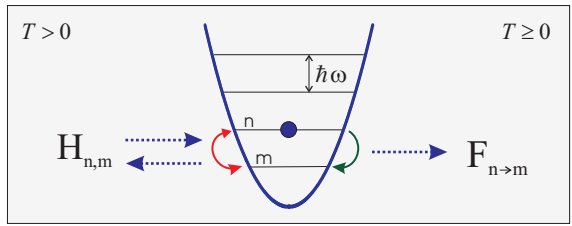}
\caption{The motion of an atom in a harmonic trap immersed in a superfluid is cooled by the generation of excitations in the superfluid. Terms in the equations of motion with coefficients $F_{n\rightarrow m}$ describe transitions from oscillator state $\ket{n}$ to state $\ket{m}$ by creation of excitations, whilst finite temperature contributions with coefficients $H_{n,m}$ account for the interaction of the system with thermal excitations.}
\label{qbitcool}
\end{figure}

A master equation is derived for the density operator of this system (section \ref{dampingeq} and appendix \ref{appmastereq}), from which the time evolution of the probability $p_n$ that the atom is in the $n$th motional state of the Harmonic oscillator potential is shown to be
\beq
\dot{p}_m=\sum_{n>m}F_{n\rightarrow m} p_n - \sum_{n^\prime<m} F_{m \rightarrow n^\prime} p_m +\sum_n H_{n,m} (p_n-p_m).
\eeq
The terms with coefficient $F_{n\rightarrow m}$,
\beq
F_{n\rightarrow m}=\frac{2\pi}{\hbar}\sum_\bfq |Z_{n,m}(\bfq)|^2 \delta(\hbar \omega (n-m) - \varepsilon_q) ,
\eeq
where $Z_{n,m}$ are the matrix elements of the interaction Hamiltonian in the basis of Harmonic oscillator energy eigenstates (Fock states), describe the transitions from state $n$ to state $m$ due to generation of excitations in the superfluid, and the terms with coefficient $H_{n,m}$,
\beq
H_{m,n}=\frac{2\pi}{\hbar}\sum_\bfq N(\bfq) |Z_{n,m}(\bfq)|^2 \delta(\hbar \omega |n-m| - \varepsilon_q), 
\eeq
describe the transitions between state $n$ and state $m$ due to interactions with thermal excitations at finite temperatures. This is illustrated in Fig.~\ref{qbitcool}.

If the speed of sound in the superfluid is $u$, and the mass of superfluid atoms is $m_b$, then the behaviour of the cooling process can be separated into two regimes - where the motion of the oscillating atom is subsonic ($\hbar \omega \ll m_b u^2/2$) or supersonic ($\hbar \omega \gg m_b u^2/2$). In the supersonic regime, cooling from any excited oscillator state occurs directly to all lower energy states, including a significant transition rate directly to the ground state (section \ref{quantsupersonic}). The resulting rate of energy loss $\dot{\varepsilon}(n)$ for a particle in the $n$th oscillator state is not linear in $n$, but instead (for lattice and superfluid atoms of equal mass $m$) is found to be
\beq
\dot{\varepsilon}(n)\approx-\frac{g_{ab}^2 \rho_0 m^{3/2}}{\pi \hbar^{4} \sqrt{2}}\alpha [\varepsilon(n)]^{3/2},
\eeq
where $g_{ab}$ is the coupling constant for interactions between the atoms in the lattice and the superfluid, $\rho_0$ is the condensate density, $\varepsilon(n)=\hbar \omega n$, and $\alpha \sim 0.3$ is a constant. 

If we consider the slowest transition rate, that from the first excited state to the ground state, we find that the characteristic transition time, $\tau$, is given by
\beq
\frac{\omega \tau_{1\rightarrow 0}}{2\pi}\sim 1.2 \times 10^{-2} \times \frac{1}{\rho_0 a_{ab}^3}\frac{a_{ab}}{l_0},
\eeq
where $a_{ab}$ is the scattering length for the interaction between atoms in the lattice and the superfluid, and $l_0=\sqrt{\hbar/(m\omega)}$ is the size of the harmonic oscillator ground state. Thus, as $l_0$ is typically an order of magnitude larger than $a_{ab}$, and $\rho_0 a_{ab}^3\sim 10^{-4}$, $\tau$ is of the order of 10 oscillator cycles. This is a sufficiently rapid cooling rate to be useful experimentally.

In the subsonic regime, damping still occurs because the oscillatory motion of the atom is accelerated (section \ref{quantsubsonic}). Significant rates are found only for transitions between neighbouring oscillator levels, and $\dot{\varepsilon}(n)$ is found to be linear in $n$,
\beq
\dot{\varepsilon}(n)\approx-\frac{g_{ab}^2 \rho_0 \omega^4}{12 \pi m_a m_b u^7}\varepsilon(n).
\eeq

When the superfluid is at a finite superfluid temperature $T$, the system is cooled to the temperature $T$ of the superfluid. The final distribution of occupation probabilities is shown to be a Boltzmann distribution,
\beq
\bar{p}_n=\bar{p}_0\rme^{-n\hbar \omega/(k_B T)}=\left(1-\rme^{-\hbar\omega/(k_B T)}\right)\rme^{-n\hbar\omega/(k_BT)},
\eeq
where $k_B$ is the Boltzmann constant (section \ref{heating}). If the temperature corresponds to an energy much smaller than the Harmonic oscillator spacing, $k_B T \ll \hbar\omega$, then the population in excited motional states is negligible. For example, if $\omega\sim 2 \pi \times 10^5$s$^{-1}$, $\hbar\omega/k_B\sim$ 5$\mu$K, so that for $T=500$nK, we then obtain $1-\bar{p}_0\approx 5\times 10^{-5}$.

A semiclassical treatment of this system in the WKB approximation (section \ref{semiclassical}) gives a result for the supersonic case which is different from the full quantum result by only $12\%$. A similar treatment in the strongly subsonic regime gives exact agreement with the earlier result.

In section \ref{onedim} we investigate a somewhat different model for the excitations, in the context of a quasi-1D superfluid. The resulting damping rates are found to be small except in the regime where the superfluid is very strongly interacting, which is a difficult regime to obtain experimentally. Finally, appendix \ref{apptermspp} contains semiclassical estimates for small additional damping terms which arise at finite temperatures and have been neglected in earlier calculations.

\section{The Model}
\label{model}
\subsection{Avoiding Decoherence}
\label{decoherencesupp}
The total Hamiltonian of the cooling process for a single atom can be written as
\beq
\hat{H}_{\rm tot}=\hat{H}_{\rm qubit} + \hat{H}_{\rm motion} +\hat{H}_{\rm superfluid} +\hat{H}_{\rm int},
\eeq
where $\hat{H}_{\rm qubit}$ is the Hamiltonian for the internal states of the atom, denoted $\ket{0}$ and $\ket{1}$, on which the qubit is encoded, $\hat{H}_{\rm motion}$ is the Hamiltonian for the atomic motion of the atom which is to be cooled, $\hat{H}_{\rm superfluid}$ is the Hamiltonian for the superfluid, and $\hat{H}_{\rm int}$ describes the interaction between the atom and the superfluid.
In order to cool a qubit without decoherence, the internal state of the atom being cooled should remain unchanged during the cooling process. If we write the initial internal state of an atom in a particular lattice site as $\ket{\psi}$, and the combined density operator for the initial mixed motional state of the atom and the state of the superfluid as $\hat{W}(0)$, so that the total initial density operator is $\ket{\psi}\bra{\psi} \otimes \hat{W}(0)$, then the overall Hamiltonian for the cooling process, $\hat{H}_{\rm tot}$, must satisfy 
\beq
\rme^{-\rmi \hat{H}_{\rm tot} t/\hbar} \ket{\psi}\bra{\psi} \otimes \hat{W}(0) \rme^{\rmi \hat{H}_{\rm tot} t/\hbar}= \ket{\psi}\bra{\psi} \otimes \hat{W}(t). \label{coherencereq}
\eeq
Thus $\hat{H}_{\rm tot}$ must be of the form $\hat{H}_{\rm tot}=\left(\ket{0}\bra{0}+\ket{1}\bra{1}\right) \otimes \hat{H}$. This requirement is satisfied provided that the interaction Hamiltonian, $\hat{H}_{\rm int}$ is independent of the internal state of the atom in the lattice. Thus, the trap potential must be the same for the two internal states $\ket{0}$ and $\ket{1}$, and the scattering length $a_{ab}$ between atoms in the superfluid and atoms in the lattice \cite{ultracoldreview} must also be the same for the two internal states. The identical scattering lengths can be arranged by choosing symmetric spin configurations, for example, by choosing $\ket{0}$ and $\ket{1}$ to be internal states with angular momentum quantum number $F=1$ and magnetic quantum numbers $m_F=\pm 1$, and the superfluid atoms to be in an internal state with $F=1$ and $m_F=0$. In order to make such a configuration stable against spin-exchanging collisions \cite{stoofverhaar}, these states should all be in the ground state of the manifold, and to prevent the creation of pairs of lattice atoms from superfluid atoms \cite{duanpair}, the energy of the $m_F=0$ level should be lowered with respect to the $m_F=\pm 1$ states (for example by using a laser \cite{sorensen1}). 
We must also ensure that when we have $N$ qubits ($N>1$), the entanglement between them is not destroyed when the motion of one or more of them is cooled. The condition in (\ref{coherencereq}) is once again sufficient for the suppression of decoherence, but now $\ket{\psi}$ is the total internal state of the $N$-qubit system, and $\hat{W}$ is the total combined density operator for the motional state of each qubit and the state of the environment. Physically, the condition is now modified so that the interaction between any atom and the superfluid must be both independent of the internal state of that atom, and independent of the internal state of all other atoms. Because the interaction is a density-density interaction, this second requirement is always fulfilled. Note that when the correlation length of the superfluid is shorter than the separation between atoms, it is possible for the motional state of different atoms to become entangled. However, this will not affect the state of the $N$-qubit system, as the qubits are encoded solely on the internal states of the atoms, which remain at all times separable from the motional states.

\subsection{Hamiltonian for the Oscillator-Superfluid Interaction}
After imposing the requirement from the previous section, we consider only the motional degrees of freedom of the atoms in the optical lattice, which are assumed to be confined in particular lattice sites where the motional states can be approximated as those of an harmonic oscillator. Coupling to the superfluid occurs in the form of a density-density interaction that generates excitations in the superfluid, which we model as Bogoliubov excitations in a weakly interacting Bose gas \cite{ll9}. The Hamiltonian for the combined system of an atom in a lattice site and the superfluid (for the motional atomic degrees of freedom only) is given by
\beq
\hat{H}=\hat{H}_{\rm motion} + \hat{H}_{\rm superfluid} +\hat{H}_{\rm int},
\eeq
where $\hat{H}_{\rm motion}$ is a 3D harmonic oscillator Hamiltonian with frequency $\omega$, which describes the motional state of the atom, $\hat{H}_{\rm superfluid}$ is the Hamiltonian for the superfluid excitations, and $\hat{H}_{\rm int}$ is the interaction Hamiltonian. 
\beq
\hat{H}_{\rm superfluid}=E_0+\sum_{\bfq \neq 0} \varepsilon(\bfq) \, \hat{b}_\bfq^\dag \hat{b}_\bfq,
\eeq
where $\hat{b}^\dag_\bfq$ and $\hat{b}_\bfq$ are creation and annihilation operators for Bogoliubov excitations in the superfluid with momentum $\hbar \bfq$ and energy $\varepsilon(\bfq)$, and $E_0$ is the ground state energy of the superfluid. 
\begin{eqnarray}
\hat{H}_{\rm int}&=&g_{ab}  \int \delta\hat{\rho}(\bfr) \,\delta\hat{\rho}_{\rm atom}(\bfr) \rmd^3\bfr \nonumber\\
&=& g_{ab}\int \delta\hat{\rho}(\bfr) \,\delta(\bfr-\hat{\bfr}) \rmd^3\bfr = g_{ab} \delta\hat{\rho}(\hat{\bfr}) \label{hint},
\end{eqnarray}
where $\delta\hat{\rho}_{\rm atom}$ is the density operator for the motion of the atom, $\hat{\bfr}$ is the position operator for the atomic motional states, $\delta\hat{\rho}$ is the density fluctuation operator in the superfluid, and $g_{ab}=4\pi\hbar^2 a_{ab}/(2 \mu)$ is the coupling constant for the interaction, with $a_{ab}$ the scattering length for interactions between superfluid atoms and atoms in the lattice \cite{ultracoldreview}, and $\mu=(m_a m_b)/(m_a+m_b)$ the reduced mass of an atom in the lattice with mass $m_a$ and a superfluid atom with mass $m_b$. 
The density fluctuation operator may be expressed as $\delta \hat{\rho}=\hat{\Psi}^\dag \hat{\Psi} - \rho_0$ where $\hat{\Psi}=\sqrt{\rho_0}+\delta \hat{\Psi}$ is the second quantised field operator for the superfluid and $\rho_0$ is the mean condensate density. In terms of the creation and annihilation operators for Bogoliubov excitations, we can write
\beq
\delta\hat{\Psi}=\frac{1}{\sqrt{V}}\sum_\bfq \left( u_{\bfq} \hat{b}_{\bfq}\rme^{\rmi \bfq.\bfr}+v_{\bfq} \hat{b}^\dag_{\bfq} \rme^{-\rmi \bfq.\bfr}\right),
\eeq
where $V$ is the normalisation volume, 
\beq
u_q=\frac{L_q}{\sqrt{1-L_q^2}}, \;\; v_q=\frac{1}{\sqrt{1-L_q^2}},
\eeq 
and 
\beq
L_q=\frac{\varepsilon_q-(\hbar q)^2/(2m)-m u^2}{m u^2}.
\eeq
The energy of excitations with momentum $\hbar \bfq$ is 
\beq
\varepsilon_q=[u^2 (\hbar q)^2 + (\hbar q)^4/(2 m_b)^2]^{1/2},
\eeq
and the speed of sound can be expressed as, $u=\sqrt{g_{bb}\rho_0/m_b}$, where $g_{bb}=4\pi\hbar a_{bb}/m_b$ with $a_{bb}$ the scattering length for interactions between atoms in the superfluid \cite{ll9}.
In a weakly interacting Bose gas at sufficiently low temperatures (where the condensate density is much smaller than the density of the normal component), the term from $\delta\hat{\Psi}^\dag \delta\hat{\Psi}$ may be neglected (see appendix \ref{apptermspp}). In this case we can write
\beq
\delta\hat{\rho}=\sqrt{\frac{\rho_0}{V}}\sum_\bfq \left( (u_\bfq+v_\bfq) \hat{b}_\bfq \rme^{\rmi \bfq.\bfr}+(u_\bfq+v_\bfq) \hat{b}^\dag_\bfq \rme^{-\rmi \bfq. \bfr}\right). \label{densfluctop}
\eeq

For the motion of atoms in the lattice, we make the approximation that the damping of in each dimension can be can be considered independently, and thus we treat the atom as a 1D oscillator with frequency $\omega$, i.e.,
\beq
\hat{H}_{\rm atom}=\hbar \omega\left(\hat{a}^\dag \hat{a} +\frac{1}{2} \right),
\eeq
where $\hat{a}$ is the lowering operator for the motional state of the atom. The position operator for the 1D oscillator is $\hat{x}=\sqrt{\hbar/(2 m_a \omega)}(\hat{a}+\hat{a}^\dag)$, where $m_a$ is the mass of the atoms in the lattice. We can also write $\bfq.\hat{\bfr} \rightarrow q_x \hat{x}$, where $q_x$ is the component of $\bfq$ in the direction of the oscillator motion.

\subsection{Damping Equations}
\label{dampingeq}
In deriving equations for the damping of the system we assume that the cooling rate is significantly slower than the period of the oscillator, and we treat the BEC as a reservoir in which the correlation time is much shorter than the correlation time of atoms in the lattice. Under these assumptions, the master equation for reduced density operator describing the motional state of an atom in the lattice, $\hat{w}(t)={\rm Tr_R}[\hat{W}(t)]$, where ${\rm Tr_R}$ denotes the trace over the superfluid states, is derived in appendix \ref{appmastereq}. 

We define the projection operator, $\hat{\mathcal{P}}$, onto a basis diagonal in the oscillator Fock states $\ket{m}$ ($H_{\rm motion} \ket{m}=\hbar\omega(m+1/2)\ket{m}$) as
\beq
\hat{\mathcal{P}}\hat{X}=\sum_m \ket{m}\bra{m}\; \bra{m} \hat{X} \ket{m},\label{projopdef},
\eeq
so that
\beq
\hat{\mathcal{P}}\hat{w}(t)= \sum_m \ket{m}\bra{m} p_m.
\eeq
Because we assumed that the oscillator trap frequency $\omega \gg \tau^{-1}$, where $\tau$ is the characteristic timescale on which transitions take place due to interaction with the superfluid, the coupling to off-diagonal elements of $\hat{w}(t)$ in the Fock state basis is very small, and the state occupation probabilities $p_n$ satisfy a closed set of equations. From appendix \ref{appmastereq} we then see that
\beq
\dot{p}_m=\sum_{n>m}F_{n\rightarrow m} p_n - \sum_{n^\prime<m} F_{m \rightarrow n^\prime} p_m +\sum_n H_{n,m} (p_n-p_m). \label{probevol}
\eeq
Here $F_{n\rightarrow m}$ is the damping coefficient at zero temperature for transitions from state $n$ to state $m$, and $H_{n,m}$ are the coefficients of the finite temperature corrections due to the absorption and scattering of thermal excitations.
The zero temperature damping coefficients are are 
\beq
F_{n\rightarrow m}=\frac{2\pi}{\hbar}\sum_\bfq |Z_{n,m}(\bfq)|^2 \delta(\hbar \omega (n-m) - \varepsilon_q)\label{fermigr} \label{dampcoef},
\eeq
and the system matrix elements of the interaction Hamiltonian are given by
\beq
Z_{n,m}(\bfq)=\bra{m}\hat{H}_{\rm int}\ket{n}=(u_\bfq+v_\bfq) \frac{g_{ab}\sqrt{\rho_0}}{V}\bra{m}\rme^{-\rmi q_x \hat{x}} \ket{n}.\label{matrixelts}
\eeq
Note that (\ref{fermigr}) is Fermi's Golden rule for the transition from state $\ket{m}$ to state $\ket{n}$ via interaction with density fluctuations in the superfluid, and that the restrictions on the summation in (\ref{probevol}) are written for clarity, but actually result from the delta function in ($\ref{fermigr}$), due to which $F_{n\rightarrow m}$ is only nonzero when $n>m$.

The finite temperature corrections are given by
\beq
H_{m,n}=\frac{2\pi}{\hbar}\sum_\bfq N(\bfq) |Z_{n,m}(\bfq)|^2 \delta(\hbar \omega |n-m| - \varepsilon_q), \label{heatcoef}
\eeq
where $N(\bfp)=(\exp[\varepsilon_p/(k_B T)]-1)^{-1}$ is the mean number of thermal Bogoliubov excitations with momentum $\hbar\bfp$ present in the superfluid, with $T$ the temperature in the normal component of the gas and $k_B$ the Boltzmann constant. Because we are interested in cooling the system to its ground state, we assume that the temperature of the superfluid is small, $k_B T \ll \hbar \omega$. This is very realistic experimentally, and in this regime we can make the approximation $H_{n,m}\approx 0$. The consequences of these terms are described in section \ref{heating}.

Also note that in the derivation of these equations we assume that the terms in $\delta\hat{\rho}$ arising from $\delta\hat{\Psi}\delta\hat{\Psi}^\dag$ may be neglected. This approximation holds when the temperature is much smaller than the critical temperature, so that the condensate density, $\rho_0$, is much larger than the density of the thermal component. The contribution from these terms at finite temperatures is estimated in appendix \ref{apptermspp}.

From the transition rates, the energy dissipation rate of an oscillator in the state $\ket{n}$ can then be calculated as
\beq
\dot{\varepsilon}(n)=\sum_{m} \hbar\omega (n-m) H_{m, n}-\sum_{m<n} \hbar\omega (n-m) F_{n\rightarrow m}\label{nenloss}.
\eeq
The total energy dissipation rate for an atom in a mixed state can be written as $\dot{\varepsilon}=\sum_{n} \hbar\omega n \dot{p}_n$, which in terms of (\ref{nenloss}) is given by $\dot{\varepsilon}=\sum_{n} \dot{\varepsilon}(n) p_n$.

\subsection{Supersonic and Subsonic Motion Regimes}
\label{modelsupersub}
We note that for typical experimental parameters in the lattice and the BEC, $\hbar \omega \gg m_b u^2/2$. For example, a Rubidium BEC with density $\rho_0\sim 10^{14}$ cm$^{-3}$ and scattering length $a_{bb}\sim 100 a_0$, where $a_0$ is the Bohr Radius, has $m_bu^2/(2\hbar)=2\pi \times 3.7\times 10^2$ s$^{-1}$, whilst typically for an atom trapped in an optical lattice, $\omega \sim 2\pi \times 10^5$ s$^{-1}$. Thus, as the maximum velocity of the atom may be estimated as $\sqrt{2 \hbar \omega/m_a}$, we see that for a typical experimental system and $m_a\sim m_b$, the atom velocities are supersonic.
In this strongly supersonic regime the requirements of energy conservation in (\ref{fermigr}) mean that even for a transition between states where $m$ and $n$ differ only by 1, the excitations are in the particle branch of the Bogoliubov excitation spectrum.  In this regime, the momentum of excitations generated in the superfluid $\hbar q\gg m_bu$ as $\hbar^2 q^2/(2m_b)\geq \hbar\omega \gg m_bu^2/2$. Hence, $\varepsilon_q\approx \hbar^2q^2/(2m_b)$, and $|u_q+v_q|^2 \approx 1$. 

If the superfluid was made sufficiently dense or strongly interacting, or the oscillator frequency $\omega$ was made sufficiently small that the motion of the oscillating atom was subsonic for all oscillator states which are initially excited, then energy conservation would cause the excitations to be in the phonon branch of the spectrum. In this regime, the momentum of excitations generated in the superfluid $\hbar q \ll m_b u$, so that $\varepsilon_q\approx \hbar u q$, and $|u_q+v_q|^2\approx\hbar k/(2 m_b u)$. 

Note that the coefficients $u_\bfq$ and $v_\bfq$ can be related to the dynamic structure factor $S(\bfk,\tilde{\omega})$ of the superfluid which is often used in relevant literature \cite{structurefactor}. In terms of the symbols used here, $S(\bfk,\tilde{\omega})=|u_k+v_k|^2$. In the same way as previously discussed, $S(\bfk,\tilde{\omega})\approx 1$ for large values of $k\gg m_b u/\hbar$, whilst for small $k$, $S(\bfk,\tilde{\omega})\propto k$. 

In the following, we treat both the supersonic and subsonic regimes. As discussed previously, the supersonic regime is the more relevant of the two in current experiments. However, the subsonic regime could be specifically engineered in experiments, and  provides an interesting comparison in terms of the physics of the damping mechanism. 

\section{Results}
\label{quantum}
The matrix elements in (\ref{matrixelts}) can be expressed in the position representation as
\beq
\bra{m}\rme^{-\rmi q_x \hat{x}} \ket{n}=\int_{-\infty}^{\infty}\rme^{-\rmi q_x x} \psi_n^*(x) \psi_m(x) \rmd x,
\eeq
where $l_0=\sqrt{\hbar/(m_a \omega)}$ is the oscillator length, $q_x$ is the component of $\bfq$ in the direction of the oscillator, $\psi_n(x)=\rme^{-x^2/(2l_0^2)}H_n(x/l_0)/\sqrt{l_0 2^n n!\sqrt{\pi}}$ is the position wavefunction for the state $\ket{n}$, and $H_n(x)$ is a Hermite Polynomial. Using the identity 
$\int_{-\infty}^{\infty}\rmd x \rme^{-(x-y)^2}H_m(x)H_n(x) \rmd x=2^n \sqrt{\pi}\,m!\,y^{n-m}L_m^{n-m}(-2y^2)$,
which assumes $m\leq n$, we can express the matrix elements (for $m<n$) as
\begin{eqnarray}
\bra{m}\rme^{-\rmi q_x \hat{x}} \ket{n}&=&\frac{m!}{n!}\rme^{l_0^2 q_x^2/4}\left(\frac{l_0^2 q_x^2}{2}\right)^{n-m}\left|L_m^{n-m}\left(\frac{l_0^2 q_x^2}{2}\right) \right|,\nonumber
\end{eqnarray}
so that 
\begin{eqnarray}
& &F_{n\rightarrow m}\nonumber\\
&=&\frac{g_{ab}^2 \rho_0}{2\pi\hbar}\frac{m!}{n!}\int_0^\infty q^2 \rmd q \delta(\hbar \omega (n-m) - \varepsilon_q) \frac{\sqrt{2}}{l_0 q}|u_q+v_q|^2\nonumber\\
&\times& \int_{-l_0 q/\sqrt{2}}^{l_0 q/\sqrt{2}}\rmd \xi \rme^{-\xi^2}\xi^{2(n-m)}\left|L_m^{n-m}\left(\xi^2\right) \right|^2.\label{finalfeq}
\end{eqnarray}
This expression can be further analysed separately in the supersonic and subsonic motion regimes, where the resulting behaviour is remarkably different. 

\subsection{Supersonic Case}
\label{quantsupersonic}
\begin{figure}[htb]
\includegraphics[width=7.5cm]{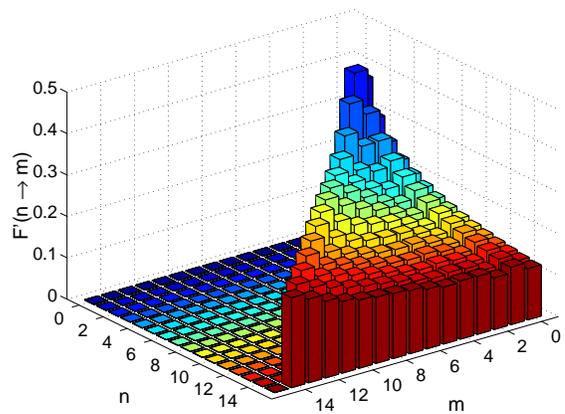}
\caption{The value of $F^\prime_{n\rightarrow m}=\pi \hbar^3 l_0 \sqrt{2}F_{n\rightarrow m}/(g_{ab}^2 \rho_0 m_b)$, showing the coefficients of the transition rate from state $n$ to state $m$ in the case of supersonic motion, as computed numerically from equation (\protect{\ref{finalsuperrate}}) with $m_a=m_b$.}
\label{super3d}
\end{figure}

Applying to (\ref{finalfeq}) the approximations given in section \ref{modelsupersub} for the case of supersonic motion yields the expression
\begin{eqnarray}
F_{n\rightarrow m}&=&\int_{-\sqrt{(n-m)(m_b/m_a)}}^{\sqrt{(n-m)(m_b/m_a)}}\rmd \xi \rme^{-\xi^2} \xi^{2(n-m)} \left|L_m^{n-m}\left(\xi^2\right) \right|^2 \nonumber\\
& &\times\frac{g_{ab}^2 \rho_0 m_b}{\pi \hbar^3 l_0 \sqrt{2}} \frac{m!}{n!}.\label{finalsuperrate}
\end{eqnarray}
The dimensionless function $F^\prime_{n\rightarrow m}=\pi \hbar^3 l_0 \sqrt{2}F_{n\rightarrow m}/(g_{ab}^2 \rho_0 m_b)$ is plotted in Fig.~\ref{super3d}, and shows the dependence of $F_{n\rightarrow m}$ on $n$ and $m$. It is immediately clear that for all $m<n$ the transition rate coefficient is significant. In fact, for all states $\ket{n}$, the transition rate directly to the ground state is of the same order as all other allowed transitions. This corresponds to the atomic motion generating a rich distribution of superfluid excitations, which is characteristic of the regime where the motion of the atom is supersonic with respect to the velocity of sound in the superfluid (see section \ref{quantsubsonic} for a comparison).

\begin{figure}[htb]
\includegraphics[width=7.5cm]{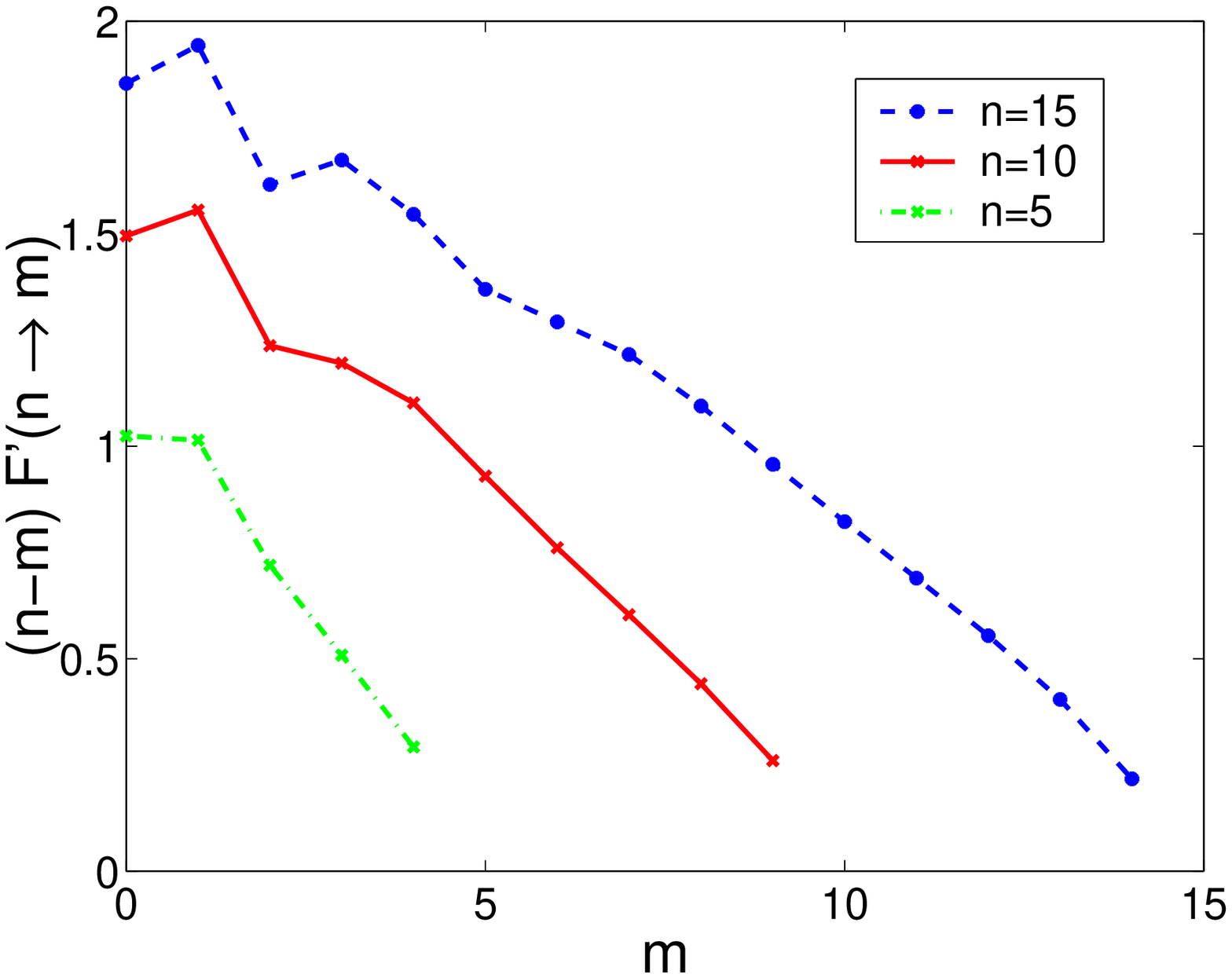}
\caption{The value of $(n-m) F^\prime_{n\rightarrow m}$, showing the contributions to the energy dissipation of the system from transitions from state $n$ to state $m$ in the case of supersonic motion. These results are computed numerically from equation (\protect{\ref{finalsuperrate}}) with $m_a=m_b$.}
\label{superm}
\end{figure}

\begin{figure}[htb]
\includegraphics[width=7.5cm]{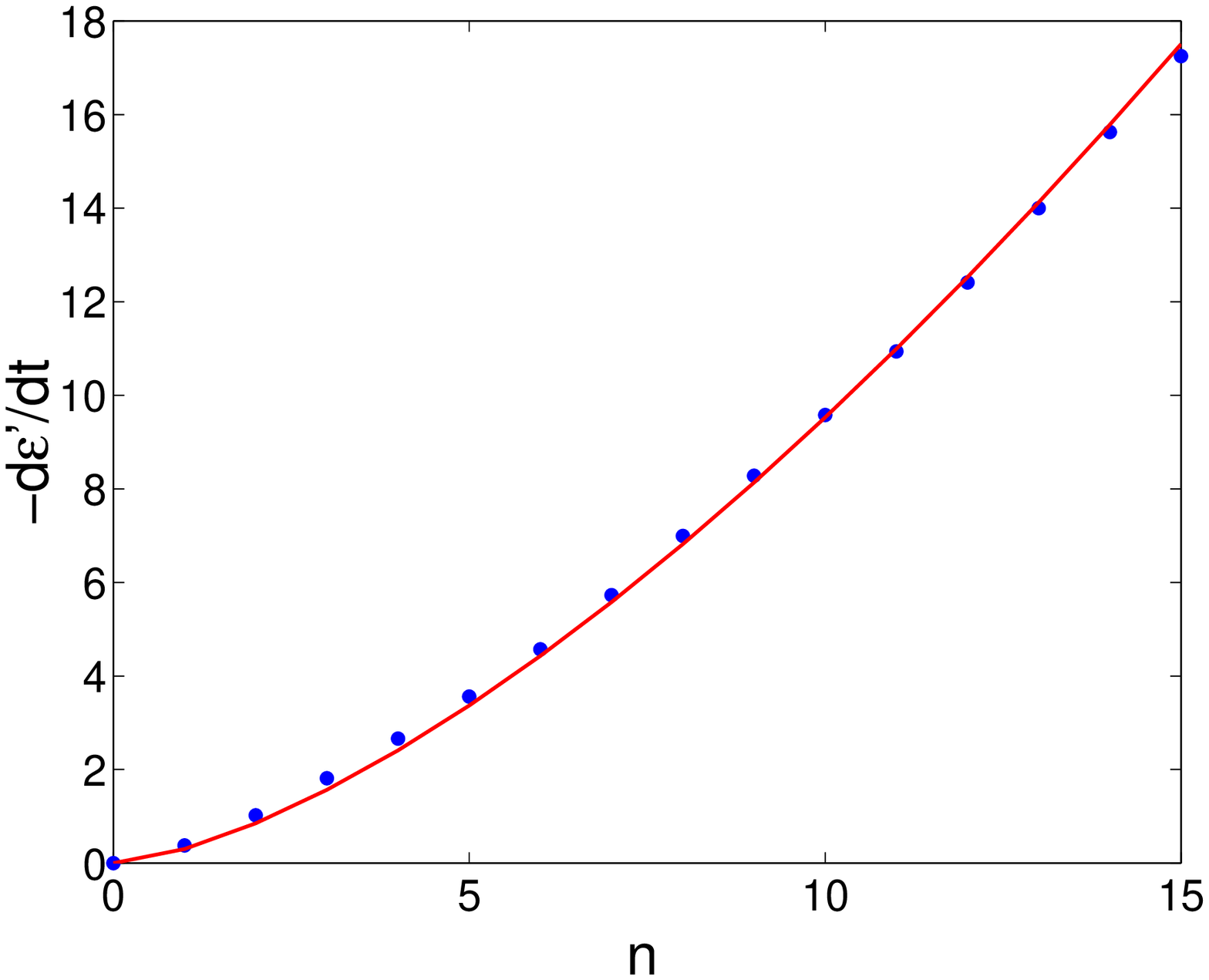}
\caption{The value of $\rmd \varepsilon^\prime/\rmd t=\sum_{(m<n)} (n-m)F^\prime_{n\rightarrow m}$ plotted as a function of $n$ for supersonic motion, showing the total rate of energy dissipation for a system instantaneously in a oscillator state with quantum number $n$. The points show the values computed numerically from equation (\protect{\ref{finalsuperrate}}) with $m_a=m_b$, and the solid line is a fitted curve of the form $\rmd \varepsilon^\prime/\rmd t=\alpha n^{3/2}$, with $\alpha=0.301$.}
\label{supern}
\end{figure}

If we consider the energy dissipation rate (\ref{nenloss}) for a system in state $\ket{n}$ in the low temperature limit ($k_BT \ll \hbar \omega$), then we see that the largest contribution comes from transitions directly to the ground state or first excited state (Fig.~\ref{superm}). In addition, the complicated excitation spectrum results in a non-exponential energy damping law, i.e., the energy dissipation rate for a state $n$ is not proportional to the energy of the state. We find instead for $m_a=m_b=m$ (Fig.~\ref{supern}) that 
\beq
\dot{\varepsilon}(n)=-\frac{g_{ab}^2 \rho_0 m^{3/2}}{\pi \hbar^{4} \sqrt{2}}\alpha [\varepsilon(n)]^{3/2},\label{finalqedot}
\eeq
where $\alpha=0.3$, and $\varepsilon(n)=\hbar \omega n$ is the energy of state $n$ measured with respect to the ground state. The total energy with respect to the ground state is then $\varepsilon= \sum_n \varepsilon(n) p_n$, so that $\dot{\varepsilon}=-\tilde{\alpha} \sum_n [\varepsilon(n)]^{3/2} p_n \approx -\tilde{\alpha} \varepsilon^{3/2}$, provided that $\varepsilon^{3/2}\approx\sum_n \varepsilon(n)^{3/2} p_n$. The time dependence of the total energy is then approximately given by
\beq
\varepsilon(t)\approx \left(\frac{1}{\varepsilon^{-1/2}(t=0) + \tilde{\alpha} t/2} \right)^2,
\eeq
where $\tilde{\alpha}=-g_{ab}^2 \rho_0 m^{3/2} \omega^{3/2}\alpha /(\pi \hbar^{5/2} \sqrt{2})$. 

The non-exponential damping law we obtain here can be understood in terms of a simple classical argument for a ``foreign'' atom moving uniformly through the superfluid at a supersonic velocity. If $\sigma_{ab}$ is the scattering cross-section for the foreign atom interacting with the superfluid, then the average number of collisions per unit time is $\rho_0 \sigma_{ab} p/m_a$, where $p/m_a$ is the velocity of the lattice atom propagating through the superfluid. The momentum of the excitation generated in a collision is $q\propto p$. Because the motion of the foreign atom is supersonic, the energy of the excitation is approximately $q^2/(2m_b)$, and the energy dissipation rate $\dot{\varepsilon}\propto \rho_0 \sigma_{ab} p^3/m_a^2 \propto \varepsilon^{3/2}$, which is the same energy dependence that we observe here.

The slowest point in the cooling process is the cooling from the first excited state to the ground state, which is also the most important case for the low energy excitations which are likely to arise in quantum computing applications. Numerically we find that for $m_a=m_b$, $F^\prime_{n\rightarrow m}=0.3789$. Thus, 
\beq
F_{1\rightarrow 0}=0.3789 \frac{g_{ab}^2 \rho_0 m}{\pi \hbar^3 l_0 \sqrt{2}}.
\eeq 
The characteristic time for the transition from the first excited state to the ground state is then expressed in terms of the number of cycles by
\begin{eqnarray}
\frac{\omega \tau_{1\rightarrow 0}}{2\pi}&=&\frac{1}{0.3789}\frac{\hbar^3 l_0 \omega}{\sqrt{2}\,  g_{ab}^2 \rho_0 m}\nonumber\\
&=&\frac{1}{0.3789} \frac{1}{16 \sqrt{2} \,\pi^2}\frac{1}{\rho_0 a_{ab}^3}\frac{a_{ab}}{l_0},
\end{eqnarray}
assuming that $m_a\approx m_b$.
In experiments, $l_0$ will typically be an order of magnitude larger than $a_{ab}$, and the parameter $\rho_0 a_{ab}^3\sim 10^{-4}$, so the characteristic transition time from the first excited state to the ground state will be of the order of 10 cycles. It is interesting that the small prefactor in this expression is very important in giving such a rapid cooling rate. This rate is sufficiently fast to be useful experimentally, particularly given that the transition rates from states with higher quantum number to the ground state are all of the same order.  

\subsection{Subsonic Case}
\label{quantsubsonic}

\begin{figure}[htb]
\includegraphics[width=7.5cm]{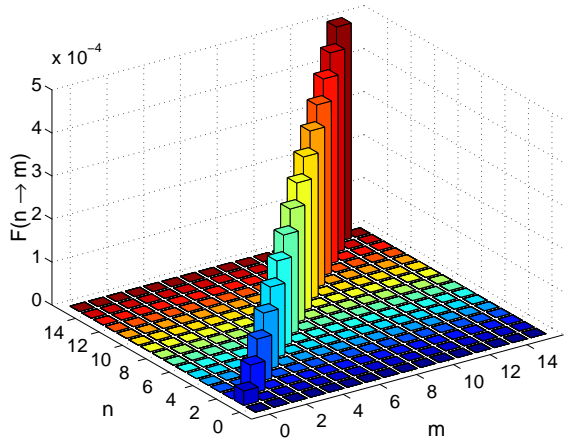}
\caption{The value of $\tilde{F}_{n\rightarrow m}=4\pi m_b \hbar u^5 F_{n\rightarrow m}/(g_{ab}^2 \rho_0 \omega^3)$, showing the coefficients of the transition rate from state $n$ to state $m$ in the case of subsonic motion, as computed numerically from equation (\protect{\ref{finalsubrate}}) for $l_0 \omega/u=0.01$.}
\label{sub3d}
\end{figure}

\begin{figure}[htb]
\includegraphics[width=7.5cm]{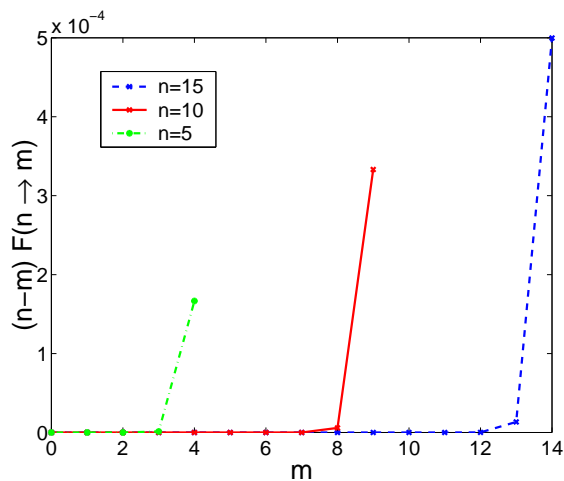}
\caption{The value of $(n-m) \tilde{F}_{n\rightarrow m}$, showing the contributions to the energy dissipation of the system from transitions from state $n$ to state $m$ in the case of subsonic motion. These results are computed numerically from equation (\protect{\ref{finalsubrate}}) for $l_0 \omega/u=0.01$.}
\label{subm}
\end{figure}

Applying the approximations given in section \ref{modelsupersub} for the subsonic case to (\ref{finalfeq}), we obtain
\begin{eqnarray}
F_{n\rightarrow m} &=& \int_{-l_0 \omega (n-m)/(u\sqrt{2})}^{l_0 \omega(n-m)/(u\sqrt{2})}\rmd \xi \rme^{-\xi^2} \xi^{2(n-m)} \left|L_m^{n-m}\left(\xi^2\right) \right|^2 \nonumber\\
& & \times  \frac{g_{ab}^2 \rho_0 \omega^3}{4 \pi m_b \hbar} \frac{m!}{n!} \frac{u\sqrt{2}}{l_0 \omega}\label{finalsubrate}.
\end{eqnarray}
Fig.~\ref{sub3d} shows $\tilde{F}_{n\rightarrow m}=4\pi m_b \hbar u^5 F_{n\rightarrow m}/(g_{ab}^2 \rho_0 \omega^3)$ plotted as a function of $n$ and $m$. In contrast to the supersonic case, we see that $F_{n\rightarrow m}$ is very sensitive to the difference $(n-m)$, and for sufficiently small $\hbar\omega/m_a u^2$, the only significant contribution to transitions from the state $\ket{n}$ are transitions to state $\ket{n-1}$. This can be seen very clearly in Fig.~\ref{subm}, which shows the contributions to the overall energy dissipation from the state $\ket{n}$.

\begin{figure}[htb]
\includegraphics[width=7.5cm]{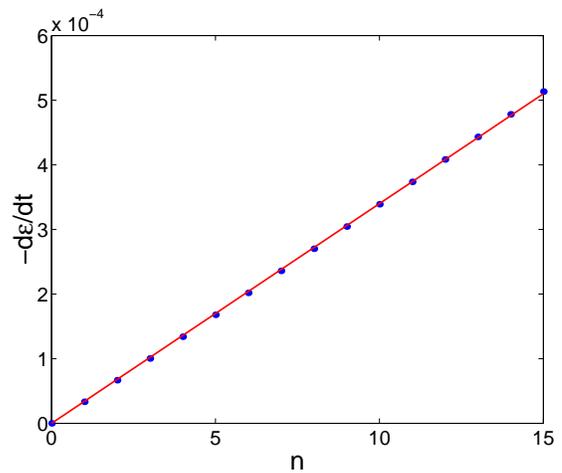}
\caption{The value of $\rmd \tilde{\varepsilon}/\rmd t=\sum_{(m<n)} (n-m)\tilde{F}_{n\rightarrow m}$ plotted as a function of $n$ for subsonic motion, showing the total rate of energy dissipation for a system instantaneously in a oscillator state with quantum number $n$. The points show the values computed numerically from equation (\protect{\ref{finalsubrate}}), and the solid line is a fitted straight line of the form $\rmd \tilde{\varepsilon}/\rmd t=\alpha \, n$, with $\alpha=3.40 \times 10^{-5}$.}
\label{subn}
\end{figure}

If we investigate the rate of energy loss $\dot{\varepsilon}(n)$ from the state $\ket{n}$ as given by (\ref{nenloss}), the only significant contribution comes from the term where $m=n-1$. For very small $\hbar\omega/m_a u^2$, we can then expand the integrand near $\xi=0$, and, noting that $L_{n-1}^1(0)=n$, obtain
\begin{eqnarray}
\dot{\varepsilon}(n)&\approx&-\frac{g_{ab}^2 \rho_0 \omega^3}{4 \pi m_b \hbar} \frac{(n-1)!}{n!} \frac{2 l_0^2 \omega^2}{6 u^2} \hbar \omega n^2 \nonumber\\
&=&-\frac{g_{ab}^2 \rho_0 \omega^4}{12 \pi m_a m_b u^7}\hbar \omega n.\label{finalresultqsub}
\end{eqnarray}
Thus, as the energy of state $\ket{n}$ measured with respect to the ground state is $\varepsilon(n)=\hbar\omega n$, the energy damping law is exponential. This is a direct consequence of the fact that, in contrast to the supersonic case, only the decay mode into the next lowest oscillator state is significant in the damping process. Fig.~\ref{subn} illustrates from numerical calculations the linear dependence of the damping rate on $n$. Note that damping still occurs in this regime despite the fact that the velocity of the atom is slower than the speed of sound in the superfluid. This apparently contradicts the Landau derivation of the critical velocity in the superfluid. However, we note that the Landau criterion is a thermodynamic argument, and cannot be applied here, as the motion is accelerated. In fact, this damping law has an analogy with that for dipole radiation in classical electrodynamics (see section \ref{subsonic}).

An analogy also exists between the supersonic and subsonic motion regimes here and regimes of large and small Lamb-Dicke parameter respectively in the context of laser cooling of trapped ions in a harmonic potential. In that system the Lamb-Dicke parameter, $\eta$, is the ratio of the size of the ground state wavefunction to the wavelength of the cooling laser, and the interaction Hamiltonian for the system is proportional to $\rme^{\rmi k \hat{x}}=\rme^{\rmi \eta (\hat{a}+\hat{a}^\dag)}$ \cite{qcompgeneral}. Here the interaction Hamiltonian is proportional to $\rme^{i q_x \hat{x}}$, and whilst $\eta$ is a fixed parameter and $q_x$ is not, $q_x$ is constrained to be a small or large parameter by the conservation of momentum when excitations are generated in the superfluid. 

During the cooling process, the coupling that exists between two states, which is proportional to $\left|\bra{m}\exp(-\rmi \eta \hat{x}) \ket{n}\right|^2$ is then analogous in the two cases. We observe cooling directly to the ground state from all excited states in the supersonic regime, and this is also a characteristic of cooling schemes in ion traps with a large Lamb-Dicke parameter. When $\eta$ or $q_x$ are small (the subsonic regime or small Lamb-Dicke parameter limit), the matrix elements simplify for $m\neq n$, $\bra{m}\exp[\rmi \eta (a+a^{\dag})] \ket{n} \approx \bra{m}\rmi \eta (\hat{a}+\hat{a}^{\dag}) \ket{n}$, and coupling only exists between nearest neighbour states (this is known in ion trap cooling as coupling to the red and blue sidebands only).

\subsection{Finite Temperature Effects}
\label{heating}

At finite temperatures, the terms proportional to $H_{n,m}$ in (\ref{probevol}) contribute heating effects due to the absorption of thermal excitations in the superfluid. When the temperature is significant, the final equilibrium motional state distribution will contain non-zero excited state probabilities. These can be calculated using the detailed balance condition \cite{gardiner}. Considering the transfer rates for atoms between oscillator states with consecutive quantum numbers, we write for the equilibrium probability distribution $\bar{p}_n=p_n(t\rightarrow \infty)$,
\beq
F_{n+1 \rightarrow n} \bar{p}_{n+1}=H_{n+1,n}(\bar{p}_{n}-\bar{p}_{n+1}),
\eeq
so that
\beq
\bar{p}_{n+1}=\frac{H_{n+1,n}}{F_{n+1\rightarrow n}+H_{n+1,n}} \bar{p}_n.
\eeq
Substituting the expressions from (\ref{dampcoef}) and (\ref{heatcoef}), and integrating over the modulus of $\bfq$, this expression simplifies to
\beq
\bar{p}_{n+1}=\frac{N(q_1)}{N(q_1)+1}\bar{p}_n,
\eeq
where $\hbar q_1=\sqrt{2 \hbar m_b \omega}$ is momentum of excitations with energy $\varepsilon_{q_1}=\hbar\omega$. Thus, $\bar{p}_{n}=[N(q_1)/(N(q_1)+1)]^n \bar{p}_0$, and using the normalisation condition $\sum_{n=0}^\infty \bar{p}_n=1$, we obtain
\beq
\bar{p}_n=\bar{p}_0\rme^{-n\hbar \omega/(k_B T)}=\left(1-\rme^{-\hbar\omega/(k_B T)}\right)\rme^{-n\hbar\omega/(k_BT)}.
\eeq
Hence, the equilibrium state occupation probabilities are simply given by the Boltzmann distribution, and the probability that an atom is in the ground motional state is $\bar{p}_0=1-\rme^{-\hbar\omega/(k_B T)}$. Provided $k_BT \ll \hbar \omega$, the absorption of thermal excitations will not significantly decrease the cooling rate, and will not prevent the cooling of essentially all of the population to the ground state. This obtainable under reasonable experimental conditions, for example, if $\omega\sim 2 \pi \times 10^5$s$^{-1}$, $\hbar\omega/k_B\sim$ 5$\mu$K, so that for $T=500$nK, we then obtain $1-\bar{p}_0\approx 5\times 10^{-5}$.

\section{The Semi-Classical Approximation}
\label{semiclassical}
\subsection{Supersonic Case}

It is interesting to compare the fully quantum calculation of the damping rates to the calculation in the semi-classical approximation. Using this approximation, the calculation is performed similarly to the calculation of damping due to radiation from an oscillating charge, which provides a useful physical analogy between the two situations. 

In this calculation we make use of the relationship between quantum matrix elements and the Fourier components of the classical trajectory of the system \cite{migdal}. Strictly speaking, this approximation is valid only when the equivalent quantum matrix elements are taken between states of large quantum number, and where the difference in the quantum numbers is small relative to the quantum numbers. We will discuss the validity of the approximation in practice at the end of the calculation. The classical trajectory of the atom in the lattice may be written in 1D as $\bfr(t)\rightarrow r_{\rm max} \cos(\omega t) \hat{\mathbf{z}}$, where $\hat{\mathbf{z}}$ is the axial unit vector along the lattice. Because the motion is periodic with period $2\pi/\omega$, the frequency spectrum of the resulting excitations will be discrete with frequencies $\omega n$ for integer $n$. Analogously to (\ref{fermigr}), we then write the rate of energy dissipation for the atom in the lattice (at zero temperature) as 
\beq
\dot{\varepsilon}=-\frac{2\pi}{\hbar}\sum_\bfq \sum_n |T_\bfq(\omega n)|^2 \delta(\hbar \omega n - \varepsilon_q) \hbar \omega n,\label{startedot}\label{edotstart}
\eeq
where 
\beq
\sum_\bfq \left| T_\bfq(\omega n)\right|^2=\sum_{N_f}\left|\frac{\omega}{2 \pi}\int_0^{2\pi/\omega}\bra{N_f} \hat{H}_{\rm int} \ket{N_i} \rme^{-\rmi \omega n t}\rmd t \right|^2, \label{tqnstart}
\eeq
with $\ket{N_f}$ the final state of the superfluid (normally a state with a particular number of excitations of momentum $\hbar\bfq$). This expression is also averaged over the initial state of the system $\ket{N_i}$, which will usually correspond to a thermal distribution of excitations.

Assuming that we are in the supersonic motion regime and applying the approximations given in section \ref{modelsupersub}, we obtain
\beq
T_q(\omega n)=\frac{g_{ab}\sqrt{\rho_0}}{\sqrt{V}}\frac{\omega}{2 \pi}\int_0^{2\pi/\omega}\rme^{-\rmi q_x r_{\rm max}\cos(t)}\rme^{-\rmi \omega n t}\,\rmd t.
\eeq
Using the identity
\beq
\left|\frac{1}{2\pi}\int_0^{2\pi}\rme^{-\rmi z \cos(\zeta)}\rme^{-\rmi n \zeta}\,\rmd \zeta\right|^2=J_n^2(z),
\eeq
where $J_n(z)$ is an ordinary Bessel Function, and integrating over the angular values of $\bfq$ in spherical coordinates then gives
\begin{eqnarray}
\dot{\varepsilon}&=&-\frac{g_{ab}^2\rho_0}{2\pi} \sum_n \int_0^\infty \rmd q\, q^2 \int_{-1}^{1}d \xi J_n^2(q r_{\rm max} \xi) \nonumber\\
& &\times \delta(\hbar \omega n -\varepsilon_q) \omega n.
\end{eqnarray}
We now integrate over $q$ to give
\beq
\dot{\varepsilon}=-\frac{g_{ab}^2\rho_0 m_b^{3/2} \omega^{3/2}}{\sqrt{2}\, \pi\hbar^{5/2}} \sum_n n^{3/2} \int_{-1}^{1}d \xi J_n^2\left(\xi a\sqrt{n} \right) \label{edoteval},
\eeq
where $a=r_{\rm max}\sqrt{2 m_b \omega/\hbar}$. We can see that many values of $n$ contribute significantly to this sum, which is analogous to the full quantum result, in which many different transitions between oscillator levels had significant coefficients $F_{n\rightarrow m}$. As noted in section \ref{quantsupersonic}, this fact arises from the the motion of the oscillating atom being faster than the speed of sound in the superfluid. This spectrum of generated excitations can be seen as being analogous to the result for electromagnetic radiation from a charge moving faster than the speed of light (in a dielectric), which can be computed semi-classically using a similar method to that used here. 

\begin{figure}[htb]
\includegraphics[width=8cm]{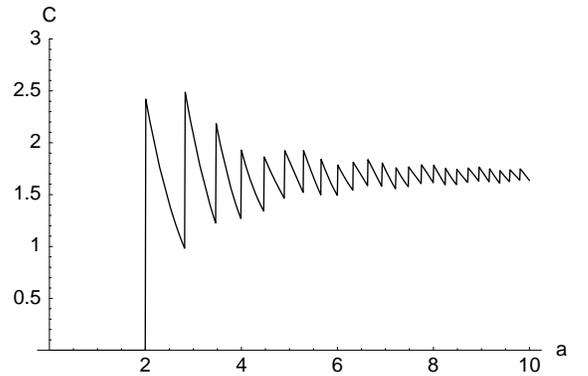}
\caption{The value of $C$ computed numerically as a function of $a=\sqrt{2}\,r_{\rm max}/l_0$ by comparison of the results from (\protect{\ref{edoteval}}) and (\protect{\ref{finaledot1}}). Note that this curve is discontinuous because of the discrete sum in (\protect{\ref{edoteval}}), which was cut off at the highest integer less than $a$, and that $C=0$ for $a<2$, because $a<2$ corresponds to a sum cut off at $n=0$. The value of this function in the limit as $a\rightarrow \infty$ gives $C\sim 1.75$.}
\label{numcomp}
\end{figure}

It is possible to determine analytically the functional dependence of (\ref{edoteval}) on $r_{\rm max}$ by finding an approximate expression for the integral over $\xi$. In the limit where the argument of the Bessel function is large, we can write
\begin{eqnarray}
F(a,n)&=&\int_{-1}^{1}d \xi J_n^2\left(\xi a \sqrt{n} \right)\nonumber\\
&\approx&2\int_{\xi_0}^{1}d \xi \frac{2 \cos^2(\xi a\sqrt{n} -n\pi/2 -\pi/4)}{\pi \xi a \sqrt{n}}\nonumber\\
&\approx&\frac{2}{\pi a \sqrt{n}}\int_{\xi_0}^1 d \xi \frac{1}{\xi}=\frac{2}{\pi a \sqrt{n}}{\rm ln}\left(\frac{a}{\sqrt{n}}\right), \label{fapprox}
\end{eqnarray}
where $\xi_0=\sqrt{n}/a$ is the lower limit for $\xi$ in which the cosine approximation of the Bessel function is valid. This expression is strictly only valid for $n\ll a^2=2 r_{\rm max}^2 m_b \omega /\hbar$. At larger values of n, $F(a,n)$ is exponentially small, and the functional dependence of $\sum_n n^{3/2} F(a,n)$ on $a$ can be found from the point at which the summation is cut off, and for a system of energy $m_a \omega r_{\rm max}^2/2$, $n_{\rm max}=m_a \omega r_{\rm max}^2/(2 \hbar)=a^2 m_a /(4 m_b)$. Approximating the sum by an integral, we can then write
\beq
\dot{\varepsilon}=-C \frac{g_{ab}^2\rho_0 m_a^2 m_b [1+2\ln(4m_b/m_a)]\,\omega^{3} r_{\rm max}^3}{32\pi^2 \hbar^{4}}. \label{finaledot1},
\eeq
where $C$ is a constant which for large values of $a$ is independent of $a$. Fig.~\ref{numcomp} shows a numerical calculation of $C(a)$, from which we observe that for large $a$, $C \sim 1.75$. Moreover, the approximation is also very good for small values of $a>2$, so that $C$ is essentially a constant for all physical values of $a$.

\begin{figure}[htb]
\includegraphics[width=8cm]{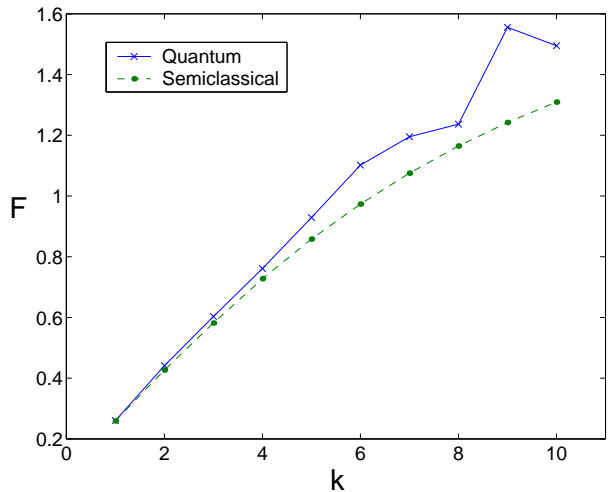}
\caption{Numerical values of the quantum result $F_{10\rightarrow(10-k)}$ (solid line) and the semi-classical result $F(20,k)$ (dotted line) in the supersonic regime. Note that we observe very good agreement for small $k$, but the results diverge for higher values of $k$.}
\label{classquant}
\end{figure}

If we use the classical expression $r_{\rm max}=\sqrt{2\varepsilon/(m_a\omega^2)}$, where $\varepsilon$ is the energy of the oscillating atom, we can rewrite (\ref{finaledot1}) as
\beq
\dot{\varepsilon}=-C [1+2\ln (4m_b/m_a)]\frac{\sqrt{2}\, g_{ab}^2\rho_0 m_a^{1/2} m_b}{16\pi^2 \hbar^4}\varepsilon^{3/2} \label{finaledot2}.
\eeq
As in the quantum case, the damping is non-exponential as a result of the rich distribution of generated excitations and instead $\dot{\varepsilon}\propto \varepsilon^{3/2}$. If we compare this result to that from equation (\ref{finalqedot}), the ratio of the semi-classical result to the quantum result for $m_a=m_b$ is $C[1+4 \ln (2)]/(8\alpha \pi)\approx 0.88$. The reason for this becomes clear when we examine the terms of the series $\sum_k k^{3/2}F(2n,k)$ (Noting that if we begin in the initial state $\ket{n}$ then $a=2 n$), and compare them to the equivalent terms in the quantum calculation, $\sum_k k F_{n\rightarrow (n-k)}$. This is shown for an initial state $n=10$ in Fig.~\ref{classquant}. We see that the terms agree well for small $k$ but that they diverge as $k\rightarrow n$. This is because the equivalence between the semi-classical result from the Fourier spectrum and the quantum matrix elements is strictly only valid when $k$ is small. Because in the calculation of energy dissipation rates the terms are weighted by an additional factor of $k$, the terms where the largest discrepancy arises are always significant in the calculation of the damping rates, and thus this discrepancy does not significantly decrease as $n\rightarrow \infty$.
 
\subsection{Subsonic Motion}

\label{subsonic}
In addition to the approximations given in section \ref{modelsupersub}, we note that for the purposes of the semi-classical calculation in the subsonic regime, $|\bfq.\bfr_{\rm max}| \leq q v_{\rm max}/\omega=v_{\rm max}/u \ll 1$.
Thus,
\begin{eqnarray}
& &\frac{1}{2 \pi}\int_0^{2\pi}\rme^{-\rmi q_x r_{\rm max}\cos\zeta}\rme^{-\rmi n \zeta}\,\rmd \zeta\nonumber\\
&\approx&\frac{1}{2 \pi}\int_0^{2\pi}q_x r_{\rm max}\cos(\zeta)\rme^{-\rmi n \zeta}\,\rmd \zeta = i \frac{q_x r_{\rm max}}{2}\delta_{n,\pm 1} , \label{smallvapprox}
\end{eqnarray}
and so
\begin{eqnarray}
\dot{\varepsilon}&=&-\frac{g_{ab}^2\rho_0}{4\pi m_b u} \int_0^\infty \rmd q\, q^3 \int_{-1}^{1}d \xi \left|\frac{q r_{\rm max} \xi}{2} \right|^2\nonumber\\
& &\times \delta(\hbar \omega -\varepsilon_q) \hbar \omega \nonumber\\
&=&\frac{-g_{ab}^2 \rho_0 \omega^4}{12 \pi u^7 m_b m_a}\,\varepsilon \label{dampratesubsonic}
\end{eqnarray}
As mentioned in section \ref{quantsubsonic}, damping occurs here despite the fact that the velocity of the atom being slower than the speed of sound in the superfluid appears to contradict the Landau derivation of the critical velocity in the superfluid, and we obtain an exponential damping law. In the same sense that the previously discussed case of supersonic motion is analogous to radiation from a charge moving faster than the speed of light in a dielectric, this case is analogous to dipole radiation from an accelerating charge. The approximation made that results in only one term in the sum being significant, (\ref{smallvapprox}), similarly corresponds to the dipole approximation in non-relativistic quantum electrodynamics. 

Note that if we substitute $\varepsilon=\hbar \omega n$ into (\ref{dampratesubsonic}), then we obtain exactly the same result we obtained from the quantum case (\ref{finalresultqsub}). The semi-classical approximation works extremely well here, because the only significant contribution to the quantum calculation comes from matrix elements between states with quantum numbers differing by one.

\section{Immersion in a Strongly Correlated 1D Superfluid}
\label{onedim}

In this section we investigate the damping that occurs when the lattice is immersed in a quasi-one dimensional superfluid, which is an example of strongly correlated quantum liquid. In a real experiment this setup is not particularly practical for cooling the motion of the atoms. For a gas to be quasi-one dimensional, the excitation modes in the transverse directions must have energies larger than all other significant energy scales in the system, and so the oscillator energies for lattice atoms, $\hbar \omega$ must be much smaller than the energies of the transverse excitations in the superfluid. Furthermore, the motion of the oscillator will only be damped in one dimension (along the direction of the quasi-1D superfluid), and so the oscillator should be made strongly anisotropic so that in the transverse directions the oscillator is always in the motional ground state and need not be cooled. However, the study of the cooling process in this context is still interesting because the lattice atom in this setup could be used as a probe to provide spectroscopic information about the 1D Bose gas.

In general the excitation spectrum of such a one-dimensional Bose gas is complicated. In the case of short-range interactions between the particles exact analytical solution exists both for the ground state wavefunction and for the excitation spectrum \cite{lieb} for arbitrary strength of the interparticle interactions and the excitation energies. However, in the limit of long wavelength the  excitations are phonons and the system can be described within a hydrodynamic approach. Following \cite{haldane} we represent the field (Bose-particle annihilation) operator in the form:  $\hat{\Psi}(x)\propto\sqrt{\rho_0+\delta{\hat{\rho}}} \rme^{\rmi \hat{\phi}}$, where $\hat{\phi}$ and $\delta\hat{\rho}$ are phase and density fluctuation fields respectively and obey the commutation relation $[\delta\hat{\rho}(x), \hat{\phi}(y)]= \rmi \delta(x-y)$, and $\rho_0$ is the 1D density (averaged, in practice, over the transverse directions). The low-energy effective Hamiltonian for the liquid is then
\beq
\hat{H}_0=\frac{\hbar}{2\pi}\int_{-\infty}^{\infty} \rmd x [v_J(\partial_x \hat{\phi})^2+v_N(\pi \delta \hat{\rho})^2],\label{hydroham}
\eeq
where $v_J=\pi \hbar \rho_0/m_b$, $v_N=\kappa/(\pi\hbar\rho_0)$,  and $\kappa$ is compressibility per unit length. The excitation spectrum corresponding to this Hamiltonian satisfies a linear dispersion relation $\varepsilon_q=\hbar v_s q$, where the velocity of sound is given by $v_s=(v_J v_N)^{1/2}$.

The parameters $v_J$ and $v_N$ are phenomenological and can be found from the exact Lieb-Liniger solution \cite{lieb}. The dependence on the interaction strength between gas particles can be described using the dimensionless parameter, $\gamma=m_b g_{bb}/(\hbar^2\rho_0)$. In the week interaction limit, $\gamma \ll 1$, the velocity of sound is given by the usual Gross-Pitaevskii value: $v_s=\sqrt{g_{bb}\rho_0/m_b}$. If the interaction is very strong, $\gamma \gg 1$, then the interaction effectively makes the particles impenetrable, and hence in a true 1D system, indistinguishable from Fermions. This is called the Tonks gas regime, and the sound velocity is equal to the effective Fermi velocity: $v_s=\pi\hbar\rho_0/m_b$. 
The energy spectrum is linear for  $\varepsilon_q \ll g_{bb}\rho_0$ (the chemical potential of a weakly interacting Bose-gas) and  $\varepsilon_q\ll \pi \hbar^2\rho_0^2/(2 m_b)$ (the Fermi energy of the Tonks gas) for the cases of weak and strong interactions, respectively. At higher energies the excitation spectrum is no longer universal and depends on the details of the interparticle interactions. Because $\varepsilon_q$ and the trapping frequency $\omega$ in the lattice are related via energy conservation, the motion of the lattice atoms must then be subsonic with respect to $v_s$ for the model to be valid.

The operator for density fluctuations in this regime is given by 
\beq
\delta\hat{\rho}=\sum_q  \left(\frac{2 q \sqrt{K}}{\pi L} \right)^{1/2}\left(\hat{b}_q \rme^{\rmi q x}+\hat{b}^\dag_q \rme^{-\rmi q x}\right), 
\eeq
where $L$ is the length of the BEC and $K=(v_J/v_N)^{1/2}$. The quantity $K$ depends on the interparticle interactions and is related to the scaling dimension of the particle field operator: $\langle \hat{\Psi}^\dag(x)\hat{\Psi}(x^\prime)\rangle\sim |x-x^\prime|^{-1/(2K)}$ for large $|x-x^\prime|$.  The function $K(\gamma)$ monotonically decreases as $\gamma$ grows, so that
$K(\gamma\rightarrow 0)\approx \pi [\gamma-(1/2\pi)\gamma^{3/2}]^{-1/2}$ and $K(\gamma\rightarrow \infty)\approx (1+2/\gamma)^2$ \cite{lieb}. Note also that for the 
quasi one-dimensional system, $g_{bb}=4 \pi \hbar^2a_{bb}/{m_b l_\perp^2}$, where $l_\perp$ is the transverse 
confinement length of the BEC, provided that $a_s \ll l_\perp$ \cite{scatter1d}. 

In the limit of small oscillation frequencies $\omega$, we apply the same approximation (\ref{smallvapprox}) used in section \ref{subsonic}, and obtain
\begin{eqnarray}
\dot{\varepsilon}&=&-\frac{2 \omega g_{ab}^2 \sqrt{K}}{\pi} \int_0^\infty \rmd q\, q \left|\frac{q r_{\rm max}}{2} \right|^2\delta(\hbar \omega -\hbar v_s q) \nonumber\\
&=&\frac{-g_{ab}^2\sqrt{K}\omega^2}{\pi\hbar m_a v_s^4}\varepsilon
\end{eqnarray}

For small $\gamma$, $K\approx\pi\hbar\sqrt{\rho_0/(m_b g_{bb})}$ and $v_s=\sqrt{g_{bb}\rho_0/m_b}$, so 
\beq
\dot{\varepsilon}\approx\frac{-g^2_{ab}\omega^2 m_b^{7/4}}{\sqrt{\pi \hbar}m_a \rho_0^{7/4}g_{bb}^{9/4}}\,\varepsilon.
\eeq
The transition rate constant is then $\Gamma_{\epsilon}\sim \omega (g_{ab}/g_{bb})^2 (\hbar \omega/\rho_0 g_{bb}) (m_b g_{bb}/\hbar^2 \rho_0)^{3/4}(m_b/m_a)/\sqrt{\pi}\ll \omega$ and hence is generally small. In the opposite limiting case for large $\gamma$, $K\approx 1$ and $v_s=\pi\hbar\rho_0/m_b$, so
\beq
\dot{\varepsilon}\approx\frac{-g^2_{ab}\omega^2 m_b^4}{\pi^5 \hbar^5 m_a \rho_0^4}\,\varepsilon.
\eeq
Here, $\Gamma_\epsilon\sim \omega (m_b g_{ab}/\hbar^2 \rho_0)^2(\omega m_b/\hbar \rho_0^2)(m_b/m_a) /\pi^5$. Thus, in this regime, the damping rates can be made very fast,
provided that $\gamma=m_b g_{ab}/\hbar^2 \rho_0$ is made very large. However, this regime is difficult to obtain experimentally, and in most current experiments $\gamma\sim 1$.

In both cases the damping that we obtain is exponential, which again arises because the motion we consider is subsonic, and produces excitations at only one significant momentum. The energy exchange rate grows as a function of $\omega$, in a manner analogous to dipole radiation in quantum electrodynamics.

\section{Summary}

We have shown that the immersion of a system of atoms in an optical lattice in a superfluid causes damping of atoms in excited motional states, and that this damping can be used to transfer these atoms to the ground motional state whilst preserving their initial internal state and any entanglement between the atoms. For typical experimental parameters, this transfer occurs in a characteristic time of around 10 oscillator cycles, which is sufficiently rapid to be useful experimentally. These typical parameters come from a regime in which the atoms in the lattice are moving faster than the velocity of sound in the superfluid, which generates a rich distribution of excitations, involving significant transitions from all levels directly to the ground state. In the opposite regime, where the velocity of the atoms in the lattice is significantly slower than the speed of sound in the superfluid, damping still occurs because the motion is accelerated, but only transitions between neighbouring oscillator levels contribute significantly to the damping process. 

Provided that the temperature in the non-superfluid fraction of the gas is much smaller than the oscillator level spacing in the lattice, heating effects due to absorption of thermal excitations is not a significant effect in this process. This is the case for experimentally realisable conditions. At higher temperatures, the system would be cooled not to the ground state, but to a thermal distribution of motional states corresponding to a Boltzmann distribution with the same temperature as that in the normal component.

The supersonic motion regime discussed here is readily realisable in present experiments. Together with a careful choice of internal atomic states used to encode a qubit, this damping mechanism thus provides a decoherence-free means to cool an atomic qubit to its motional ground state.

\begin{acknowledgments}
This work was supported in part by the Austrian Science Foundation FWF, E.U. Networks and the Institute for Quantum Information. 
\end{acknowledgments}

\begin{appendix}

\section{Derivation of the Master Equation}
\label{appmastereq}
We treat the superfluid with Bogoliubov excitations as a reservoir, with density operator $\hat{R}$. In the interaction picture, and after making the Born-Markov approximation, the master equation for the density operator $\hat{w}$ of a system which interacts with a reservoir via an interaction Hamiltonian $\hat{H}_{\rm int}$ can be shown to be given by 
\cite{wallsmilburn}
\beq
\dot{\hat{w}}=-\frac{1}{\hbar^2}\int_0^t \rmd t^\prime {\rm Tr}_R[\hat{H}_{\rm int}(t),[\hat{H}_{\rm int}(t^\prime),\hat{w}(t) \otimes \hat{R}]],\label{startmeq}
\eeq
where ${\rm Tr}_R$ denotes the trace over the reservoir states. 

We write $\hat{s}_{\bfq,1}=\rme^{\rmi q_x \hat{x}}$, $\hat{s}_{\bfq,2}=\rme^{-\rmi q_x \hat{x}}$, $\hat{\Gamma}_{\bfq,1}=\hat{b}_\bfq$ and $\hat{\Gamma}_{\bfq,2}=\hat{b}_\bfq^\dag$, so that
\beq
\hat{H}_{\rm int}=\frac{g_{ab} \sqrt{\rho_0}}{V}(u_q+v_q) \sum_\bfq \sum_{i=1,2} \hat{s}_i \hat{\Gamma}_i,
\eeq
and then substitute this expression into (\ref{startmeq}) to give
\begin{eqnarray}
\dot{\hat{w}}&=&-\frac{g^2_{ab} \rho_0 (u_q+v_q)^2}{V \hbar^2}\sum_\bfq\sum_{i,j=\{1,2\}}\int_0^t \rmd t^\prime \nonumber\\
& &\left[\hat{s}_i(t)\hat{s}_j(t^\prime)\hat{w}(t)-\hat{s}_j(t^\prime)\hat{w}(t)\hat{s}_i(t)\right]\langle\hat{\Gamma}_i(t) \hat{\Gamma}_j(t^\prime)\rangle_R \nonumber\\
&+&\left[\hat{w}(t)\hat{s}_j(t^\prime)\hat{s}_i(t)-\hat{s}_i(t)\hat{w}(t)\hat{s}_j(t^\prime)\right]\langle\hat{\Gamma}_j(t^\prime) \hat{\Gamma}_i(t)\rangle_R \nonumber,\\
\end{eqnarray}
where we have used the cyclic property of the trace, dropped the operator subscript $\bfq$, and written ${\rm Tr}_R(\hat{R}\hat{A})=\langle\hat{A}\rangle_R$. We have also used the fact that $\langle\hat{\Gamma}_{\bfq,i}(t^\prime) \hat{\Gamma}_{\bfq^\prime,j}(t)\rangle_R=0$ for $\bfq \neq \bfq^\prime$.

Proceeding in the standard way, we change the variable of integration to $\tau=t-t^\prime$, and note that $\hat{b}_\bfq(t-\tau)=\rme^{-\rmi \hat{H}_b \tau/\hbar}\hat{b}_\bfq(t)\rme^{\rmi \hat{H}_b \tau/\hbar}=\rme^{\rmi\varepsilon_\bfq \tau/\hbar}\hat{b_\bfq}(t)$ 
and similarly $\rme^{-\rmi \hat{H}_b \tau/\hbar}\hat{b}^\dag_\bfq(t)\rme^{\rmi \hat{H}_b \tau/\hbar}=\rme^{-\rmi \varepsilon_\bfq \tau/\hbar}\hat{b}^\dag_\bfq(t)$. 

We then make use of the assumption that $\hat{w}(t)\approx \hat{\mathcal{P}} \hat{w}(t)$ (see (\ref{projopdef})), and write the master equation in a Fock state representation. Noting also that $\langle \hat{b}_{\bfq}(t) \hat{b}_{\bfq}(t)\rangle_R=\langle \hat{b}^\dag_{\bfq}(t) \hat{b}^\dag_{\bfq}(t)\rangle_R=0$, we obtain

\begin{widetext}
\begin{eqnarray}
\dot{\hat{w}}&=&-\frac{2 g^2_{ab} \rho_0 (u_q+v_q)^2}{V \hbar^2}\sum_\bfq\sum_{m,n} \int_0^t \rmd \tau\left[ \ket{m}\bra{m}\rme^{-\rmi q_x \hat{x}}\ket{n}\bra{n}\rme^{\rmi q_x\hat{x}}\ket{m}\bra{m} p_m \rme^{\rmi \omega \tau (m-n)}\right.\nonumber\\
& &\left.+\ket{m}\bra{m}\rme^{-\rmi q_x\hat{x}}\ket{n}\bra{n}\rme^{-\rmi q_x.\hat{x}}\ket{m}\bra{m} p_n \rme^{\rmi \omega \tau (n-m)}\right] \left( \rme^{-\rmi \varepsilon_q \tau/\hbar}\langle\hat{b}_\bfq \hat{b}_\bfq^\dag \rangle_R +\rme^{\rmi \varepsilon_q \tau/\hbar}\langle\hat{b}_\bfq^\dag \hat{b}_\bfq \rangle_R \right).
\end{eqnarray}

Assuming that the correlation time of the superfluid reservoir is much shorter than that in the system we can extend the integration over $\tau\rightarrow \infty$, and making the replacement $\int_0^\infty\rmd \tau \rme^{\rmi (\varepsilon-\varepsilon_0) \tau/\hbar}\rightarrow \pi \hbar \delta(\varepsilon-\varepsilon_0)$, we obtain
\begin{eqnarray}
\dot{p}_m&=&\frac{2\pi g^2_{ab} \rho_0 (u_q+v_q)^2}{V \hbar}\sum_\bfq \sum_n \left|\bra{m}\rme^{-\rmi q_x \hat{x}}\ket{n}\right|^2 \left\{ \left[\delta\left(\hbar\omega(n-m) -\varepsilon_\bfq\right)p_n- \delta\left(\hbar\omega(m-n)-\varepsilon_\bfq\right)p_m \right]\langle\hat{b}_\bfq \hat{b}_\bfq^\dag \rangle_R \right.\nonumber\\
& &\left.+ \left[ \delta\left(\hbar\omega(m-n)-\varepsilon_\bfq\right)p_n - \delta\left(\hbar\omega(n-m)-\varepsilon_\bfq\right)p_m \right]\langle\hat{b}^\dag_\bfq \hat{b}_\bfq \rangle_R \right\}.\label{finalmastereq}
\end{eqnarray}
\end{widetext} 
The first two terms here (those proportional to $\langle\hat{b}_\bfq \hat{b}_\bfq^\dag \rangle_R$) describe the damping by creation of excitations in the superfluid, whilst the second two terms (those proportional to $\langle\hat{b}^\dag_\bfq \hat{b}_\bfq \rangle_R$) describe heating effects by absorption of thermally generated excitations. At finite temperatures, the reservoir correlation functions are given by the number of thermal excitations $N(\bfq)$ with momentum $\hbar \bfq$, $\langle\hat{b}^\dag_\bfq \hat{b}_\bfq \rangle_R=N(\bfq)$.

\section{Estimation of $\delta\hat{\Psi}^\dag \delta\hat{\Psi}$ Terms}
\label{apptermspp}
The heating effects due to absorption of thermal excitations has already been discussed in section \ref{heating}, where the equilibrium distribution at finite temperatures was shown to be a Boltzmann distribution. The small additional damping terms arising at finite temperatures from the $\delta\hat{\Psi}^\dag \delta\hat{\Psi}$ term, which are small when the condensate density, $\rho_0$ is large and which were omitted when the density fluctuation operator $\delta \hat{\rho}$ was originally written, may be estimated using a semi-classical treatment. The operator for the additional density fluctuation terms is given by
\begin{eqnarray}
\delta \hat{\rho}^\prime&=&\delta\hat{\Psi}^\dag \delta\hat{\Psi}=\frac{1}{V}\sum_{\bfp,\bfp'} u_\bfp u_{\bfp'} \hat{a}_\bfp \hat{a}^\dag_{\bfp'} \rme^{\rmi(\bfp-\bfp').\bfr}\nonumber\\
&+& v_\bfp v_{\bfp'} \hat{a}^\dag_\bfp \hat{a}_{\bfp'} \rme^{-\rmi(\bfp-\bfp').\bfr}+u_\bfp v_{\bfp'} \hat{a}_\bfp \hat{a}_{\bfp'} \rme^{\rmi(\bfp+\bfp').\bfr}\nonumber\\
&+& u_\bfp v_{\bfp'} \hat{a}^\dag_\bfp \hat{a}^\dag_{\bfp'} \rme^{-\rmi(\bfp+\bfp').\bfr}.
\end{eqnarray}
The first two terms in this expression correspond to the inelastic scattering of thermal excitations with momentum $\hbar\bfp$ to excitations with momentum $\hbar \bfp'$, and the second two correspond to the absorption and emission respectively of two excitations with momenta $\hbar \bfp$ and $\hbar \bfp'$. 

For the case of supersonic motion where $u_q\rightarrow 1$ and $v_q \rightarrow 0$, the correction to the dissipation rate is then given by
\begin{eqnarray}
\dot{\varepsilon}'&=&-\frac{\pi g_{ab}^2}{\hbar}\sum_{\bfp,\bfp'} \sum_n [N(\bfp)-N(\bfp')] \,\delta(\hbar \omega n - \varepsilon_{p'}+\varepsilon_{p}) \nonumber\\
& & \times \left|\frac{\omega}{2 \pi}\int_0^{2\pi/\omega} \rme^{\rmi(\bfp-\bfp').\bfr(t)} \rmd t\right|^2 \, \hbar \omega n, 
\end{eqnarray}
where, as before, $N(\bfp)=(\exp[\varepsilon_p/(k_B T)]-1)^{-1}$ is the mean number of thermal Bogoliubov excitations with momentum $\hbar \bfp$ present in the superfluid. 

In order to cool the system to the ground state we already require $\hbar \omega \gg k_B T$, which has been shown to be a reasonable experimental condition in section \ref{heating}. In this case, the thermally generated excitations with momentum $\hbar\bfp$ will have a much smaller energy than the scattered excitations, which have momentum $\hbar\bfp'$. Also, $\varepsilon_{\bfp'}>\hbar \omega \gg k_B T$ and $N(\bfp')\approx 0$. Thus,
\begin{eqnarray}
\dot{\varepsilon}'&\approx&-\frac{\pi g_{ab}^2}{\hbar}\sum_{\bfp,\bfp'} \sum_n N(\bfp) \,\delta(\hbar \omega n - \varepsilon_{p'}) \nonumber\\
& & \times \left|\frac{\omega}{2 \pi}\int_0^{2\pi/\omega} \rme^{\rmi(\bfp').\bfr(t)} \rmd t\right|^2 \, \hbar \omega n, \nonumber\\
&=& \frac{\dot{\varepsilon}}{2\rho_0} \frac{1}{2\pi^2}\int_0^{\infty}p^2 \rmd p N(\bfp).
\end{eqnarray}
This result is proportional to the density of thermal excitations and essentially describes the classical friction due to scattering of thermal excitations 
by the moving particle.

If $k_B T \ll m_b u^2/2$, then $\varepsilon_{\bfp}\approx \hbar u p$. The additional damping is then given in terms of the rate $\dot{\varepsilon}$ in (\ref{finaledot2}) by 
\beq
\dot{\varepsilon}'=\frac{\dot{\varepsilon}\zeta(3)}{2\pi^2 \rho_0 (\hbar u)^3} (k_B T)^3 , \label{tcorr1}
\eeq
where $\zeta(x)$ denotes the Riemann Zeta function. Note that because the wavenumber of phonons in this regime is of the order of $k_B T/ (\hbar u)$, this result is proportional to the density of thermal phonons, $\rho_{\rm phonons}$. Thus the additional damping term is equal to that in (\ref{finaledot2}), but with the numerical coefficient modified, and the density of the condensate $\rho_0$ replaced by the density of thermal phonons, $\rho_{\rm phonons}$. This term will always be small, as in this regime $T<T_c$, the critical temperature of the Bose gas, so $\rho_0 \gg \rho_{\rm phonons}$.

If $k_B T \gg m_b u^2/2$, then $\varepsilon_{\bfp}\approx \hbar^2 p^2/(2 m_b)$. The rate of additional damping is then
\beq
\dot{\varepsilon}'=\frac{\dot{\varepsilon}\zeta(3/2) m_b^{3/2}}{4 \sqrt{2}\, \pi^{3/2} \rho_0 \hbar^3} (k_B T)^{3/2}. \label{tcorr2}
\eeq
For a uniform Bose gas the critical temperature for Bose condensation can be expressed as \cite{pethicksmith} 
\beq
k_B T_c=\frac{2 \pi \hbar^2 \rho_t^{2/3}}{m_b [\zeta(3/2)]^{2/3}},
\eeq
where $\rho_t=\rho_0+\rho_n$ is the total density, and $\rho_n$ is the density of the normal component, so that we can rewrite (\ref{tcorr2}) as
\beq
\dot{\varepsilon}'=\frac{\dot{\varepsilon}\rho_t}{2 \rho_0} \left(\frac{T}{T_c}\right)^{3/2} = \frac{\dot{\varepsilon}\rho_n}{2 \rho_0}, 
\eeq
where we have used the well known result $\rho_n=\rho_t (T/T_c)^{3/2}$ \cite{pethicksmith}. Thus, this result has the same form as the damping rate obtained in (\ref{finaledot2}), but the condensate density is replaced by the density of the normal component, and the numerical coefficient is decreased by a factor of 2. Again, at small temperatures compared with the critical temperature, $T \ll T_c$, when $\rho_n \ll \rho_0$, the contribution from this term will be small.

The same calculation can be performed for the subsonic case. In this regime, the contribution from the terms involving $\hat{a}_\bfp\hat{a}_{\bfp'}$ and $\hat{a}^\dag_\bfp\hat{a}^\dag_{\bfp'}$ is small, because the double summation over $\bfp$ and $\bfp'$ is restricted by energy conservation such that $|\varepsilon_p +\varepsilon_{p'}|=\hbar \omega n$, and in the subsonic case, this quantity is always small. 
With respect to the subsonic energy dissipation rate in (\ref{dampratesubsonic}), $\dot{\varepsilon}$, we obtain
\beq
\dot{\varepsilon}'\approx-\frac{\pi^2 \dot{\varepsilon}}{480 \rho_0 m_b u^5 \hbar^3} (k_B T)^4.\label{tcorrsub1}
\eeq 
Note that as $\hbar \omega n \ll m_b u^2/2$, this expression is derived considering only the case where $k_B T \ll m_b u^2/2$. It can be shown that in the limit $k_B T \ll m u^2/2$ that the density of the normal component $\rho_n$ is given by \cite{pethicksmith}
\beq
\rho_n=\frac{2 \pi^2 (k_B T)^4}{45 m_b \hbar^3 u^5},
\eeq
so that we can write (\ref{tcorrsub1}) as
\beq
\dot{\varepsilon}'\approx-\frac{3 \dot{\varepsilon} \rho_n}{64 \rho_0}.
\eeq 
Again, this result is a modification of the zero-temperature damping result, with the condensate density replaced by the density of the normal component and the numerical coefficient decreased. In the limit $T\ll T_c$, as with the supersonic results, this result will be small, as $\rho_n \ll \rho_0$.

\end{appendix}

\end{document}